\documentclass[journal]{vgtc}              

\onlineid{1568}

\vgtccategory{Research}

\vgtcpapertype{system}

\title{ChatVis: Assisting and Evaluating Large Language Models for Generating Scientific Visualizations}

\author{
  Tom Peterka,
  Tanwi Mallick,
  Orcun Yildiz,
  David Lenz,
  Cory Quammen, and
  Berk Geveci
}

\authorfooter{
  \item
    Tom Peterka, Tanwi Mallick, Orcun Yildiz, and David Lenz are with the Mathematics and Computer Science Division of
    Argonne National Laboratory and contributed equally to this work.
    E-mail: \{tpeterka $|$ tmallick $|$ oyildiz $|$ dlenz\}@anl.gov
  \item
    Cory Quammen and Berk Geveci are with Kitware, Inc.
    E-mail: \{cory.quammen $|$ berk.geveci\}@kitware.com
}

\abstract{
Large language models (LLMs) are rapidly increasing in capability, but they still struggle with highly specialized programming tasks such as scientific visualization.
We present an LLM assistant, ChatVis, that aids the LLM to generate Python code for ParaView scientific visualization tasks, without the need for retraining or fine-tuning the LLM.
ChatVis employs chain-of-thought prompt simplification, retrieval-augmented prompt generation using a vector database of documentation and code examples, and error checking with iterative prompt feedback to correct errors until a visualization is produced.
An integral part of our approach is a benchmark suite of canonical visualization tasks, ParaView regression tests, and scientific use cases that includes comprehensive evaluation metrics.
We evaluate our visualization assistant by comparing  results with a variety of top-performing unassisted LLMs.
We find that all the metrics are significantly improved with ChatVis.
}


\teaser{
  \centering
  \includegraphics[width=0.65\linewidth]{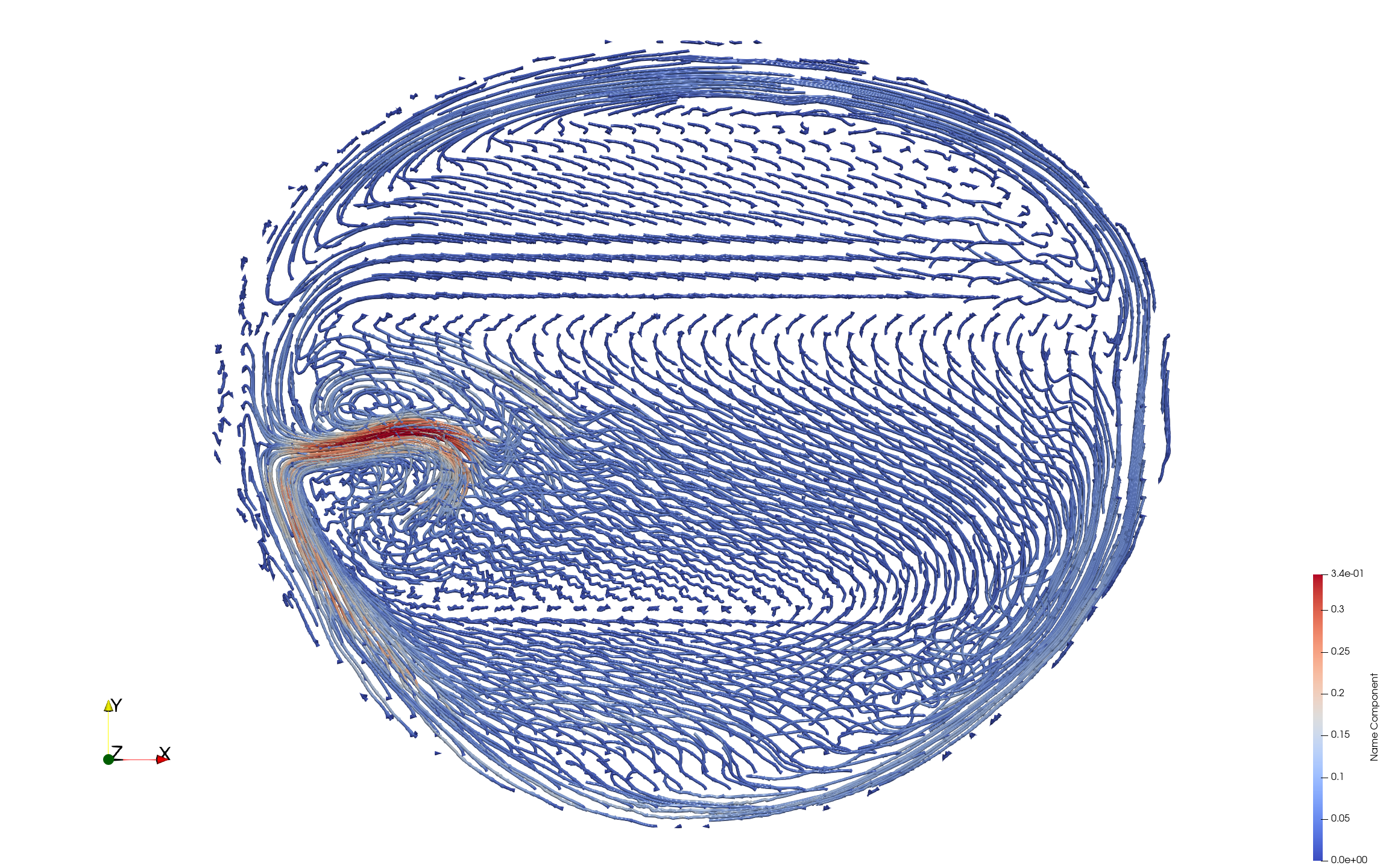}
  \caption{ChatVis-generated visualization of pathlines traced from a vector field simulated by an ocean climate model.}
  \label{fig:teaser}
}

\newcommand{\bi}{\begin{itemize}}
\newcommand{\ei}{\end{itemize}}
\newcommand{\be}{\begin{enumerate}}
\newcommand{\ee}{\end{enumerate}}
\newcommand{\im}{\item}

\usepackage{soul}

\newcommand{\kw}[1]             {{\tt #1}}

\graphicspath{{figs/}{figures/}{pictures/}{images/}{./}} 

\usepackage{mathptmx}                  
\usepackage{amsmath}

\usepackage{mdframed}
\newenvironment{myenv}[1]
  {\mdfsetup{
    frametitle={\colorbox{white}{\space#1\space}},
    innertopmargin=-3pt,
    frametitleaboveskip=-\ht\strutbox,
    frametitlealignment=\center
    }
  \begin{mdframed}%
  }
  {\end{mdframed}}

\usepackage{subcaption}

\usepackage{listings}
\usepackage{stackengine}

\usepackage{array}

\begin{document}

\firstsection{Introduction}

\maketitle

Despite decades of research and development of high-quality production software tools, 3d volumetric and 4d time-varying scientific visualization can still be daunting for nonexperts.
Besides requiring extensive knowledge of a visualization tool's interactive commands and/or application programming interface (API), users of scientific visualization require background in meshing, analytical geometry, and an awareness of the computer science algorithms underpinning fundamental data transformations and rendering operations.
Domain scientists, while expert in their respective areas such as climate science, may not have the training and experience in all of the above areas, requiring a visualization expert to assist in the use of visualization tools such as ParaView or VisIt, especially in their offline Python scripting capabilities.

With the growing ubiquity of AI assistants in many fields, and specifically with the rise of large language models (LLMs) to automate many scientific tasks, it is natural to ask whether scientific visualization can be assisted with a natural language interface supported by an LLM.
An early investigation~\cite{mallick2024chatvis} showed that off-the-shelf LLM models (ChatGPT and others) were not adequately trained in the nuances of scientific visualization, but with accompanying software to iteratively assist in prompting, could deliver promising results.

In this paper, we redesign the LLM assistant called ChatVis, where the user describes a set of scientific analysis and visualization operations in natural language, and ChatVis generates a ParaView Python script for the desired sequence of steps.
We build a vector database of ParaView documentation and consult the database using retrieval-augmented generation (RAG) for required information.
We also develop a comprehensive collection of case studies, and using these test cases we quantitatively evaluate ChatVis over a variety of metrics.
We evaluate the latest available LLMs, unaided by ChatVis, and compare performance with ChatVis.
While we use ParaView~\cite{ayachit15} as the visualization tool for this study, we believe that the Python scripting capability in VisIt~\cite{childs12} would work equally well if ChatVis were provided comparable information.

Our contributions are:
\bi
\im An LLM assistant for automated generation of scientific visualization through the Python API of a visualization tool such as ParaView
\im A methodology for collecting software documentation in a vector database and injecting it into the prompt using RAG techniques
\im A benchmark suite consisting of canonical visualization tasks, regression tests, and scientific use cases and evaluation metrics
\im Comparison of open- and closed-source unassisted LLM models with ChatVis
\im Comparison of few-shot prompting vs. RAG for visualization tasks
\im Evaluation of prompt sensitivity using two versions---longer and shorter---of user prompts

\ei

The remainder of this paper is organized as follows.
Section~\ref{sec:background} presents related work in synthetic software generation and benchmarking.
This is followed by a description of the ChatVis architecture in Section~\ref{sec:method}. Experiments are described and results tabulated in Section~\ref{sec:results}.
Section~\ref{sec:science} highlights two science use cases in greater detail that form part of our benchmark suite.
The paper concludes with a recap and look to the future of LLMs for scientific visualization in Section~\ref{sec:conclusion}.


\section{Background and Related Work}
\label{sec:background}

We survey related work in synthetic software generation using LLMs, followed by LLMs used specifically for scientific visualization.
Because we build our LLM assistant using RAG, we also include references to RAG techniques. Another key part of our work is validating correctness, so numerous related work is included in benchmarking and comparing images, videos, and code.

\subsection{Synthetic Software Generation using LLMs}

Jiang et al. survey the evolution of code generation using LLMs from 2020 to 2024~\cite{jiang2024survey}.
In the realm of software generation for HPC applications, LM4HPC by Chen et al.~\cite{chen2023lm4hpc} provides a framework for executing HPC-related coding tasks. HPC-GPT by Ding et al.~\cite{ding2023hpc} is designed for answering questions about HPC code.
 HPC-Coder by Nichols et al.~\cite{nichols2024hpc} fine-tunes a pretrained LLM to model parallel programs and accelerate developer productivity.
For Python coding, Chen et al.~\cite{chen2021evaluating} introduce Codex, a fine-tuned LLM for synthesizing programs from docstrings.

\subsection{LLMs in Visualization}

There is growing recognition of the role of LLMs in visualization.
The LLM4Vis Workshop by Hoque at IEEE VIS 2024~\cite{hoque24} identified
several applications of LLMs in visualization: visual text analytics, natural language interfaces for visualizations, text
generations for visualizations, visualization generation, automated data-driven story generation, and visualization
deconstruction and restyling. Our focus is visualization generation.

Most of the LLM work in visualization has been in visual analytics or information visualization using 2d charts.
Chart-to-text~\cite{obeid2020chart} generated natural language explanations of 2d charts using a custom deep neural network based on a transformer model.
DataNarrative~\cite{islam2024datanarrative} consists of two agents---a generator and an evaluator---assisting an LLM in the generation of narratives from data tables.
Rather than explaining existing visualizations, ChartGPT~\cite{tian2024chartgpt} generates new 2d charts from natural language prompts using an LLM fine-tuned with example charts and prompts about them.

One of the few examples of LLMs for 3d scientific visualization is the generation of transfer functions for direct volume rendering using natural language~\cite{jeong2024text}.
We believe our work is one of the first to explore LLMs in 3d and 4d (time-varying) scientific visualization generation.
Our work builds on a short workshop proof-of-concept paper by Mallick et al.~\cite{mallick2024chatvis}.
The implementation in that paper was based on few-shot prompting (not RAG); only 5 canonical visualizations were tested, without metrics for the resulting image quality, with only one version of user prompt, and comparisons were with LLM models that were available at that time.

\subsection{Retrieval-Augmented Generation (RAG)}

Retrieval-augmented generation, or RAG, is a technique for integrating external information not included in an LLM's training corpus into the context of an LLM query. Gao et al.~\cite{gao2023retrieval} surveys the past four years of progress in this area.

The concept of RAG was first introduced by Lewis et al.~\cite{lewis2020retrieval}, demonstrating how retrieval-augmented techniques improve the factual accuracy of language models by incorporating external knowledge sources.
For code completion, ReACC by Lu et al.~\cite{lu2022reacc} introduces a hybrid approach that combines lexical copying with semantic referencing, achieving state-of-the-art performance on the CodeXGLUE benchmark~\cite{lu2021codexglue}.
Similarly, Zhou et al. \cite{zhou2022docprompting} proposes DocPrompting, a technique that explicitly retrieves relevant documentation to improve functional correctness in code generation.
Zhang et al.~\cite{zhang2023repocoder} developed RepoCoder, which retrieves code analogies across different files within a repository to enhance code completion.
A key advancement in retrieval-augmented code generation is the integration of graph neural networks (GNNs) to retrieve semantically relevant code examples, as explored by Liu et al.~\cite{liu2020retrieval}.
Parvez et al.~\cite{parvez2021retrieval} also leverages retrieval from a large-scale source-code database to refine the generation process, ensuring that LLMs reference high-quality, contextually appropriate code snippets.

Taking retrieval augmentation a step further, Su et al.~\cite{su2024arks} introduces Knowledge Soup, a framework that aggregates multiple sources—--including web search, documentation, execution feedback, and evolved code snippets—--to iteratively refine retrieved information and improve generation quality.
OpenCodeInterpreter~\cite{zheng2024opencodeinterpreter} also demonstrates how execution capabilities and iterative refinement can further elevate open-source code generation, achieving performance levels that closely rival proprietary models like GPT-4 Code Interpreter.

\subsection{Benchmarks}

Jiang et al. list 36 benchmarks grouped by category in their survey paper~\cite{jiang2024survey}.
Additional relevant benchmarks are the following.
SciCode~\cite{tian2024scicode} is a coding benchmark curated by scientists in the fields of mathematics, physics, chemistry, biology, and materials science, with a high degree of difficulty.
In the field of visualization, evaluation and benchmarking focus primarily on 2d charts. VisEval~\cite{chen2024viseval} includes a custom dataset for multiple chart types, ground truth plotted data, and evaluators for validity, correctness, and readability.
The chart-to-text benchmark for chart summarization~\cite{kantharaj2022chart} features charts of various types and is used to evaluate several image captioning and data-to-text models.
The ChartQA~\cite{masry2022chartqa} benchmark evaluates complex reasoning about charts involving logic and arithmetic operations that refer to the visual features of a chart.

\subsection{Correctness Metrics}

The metrics we considered as candidates for our evaluation include image comparison metrics for single visualization scenes, video comparison metrics for time-varying animations, and code comparison metrics for the generated Python code.
Ultimately we found the code comparisons unreliable for our purposes, and we decided to evaluate animations using key frames of image comparison metrics. We include all three categories---image, video, and code comparison---of metrics here for completeness.

\begin{figure*}[t!]
    \centering
    \includegraphics[width=0.7\textwidth]{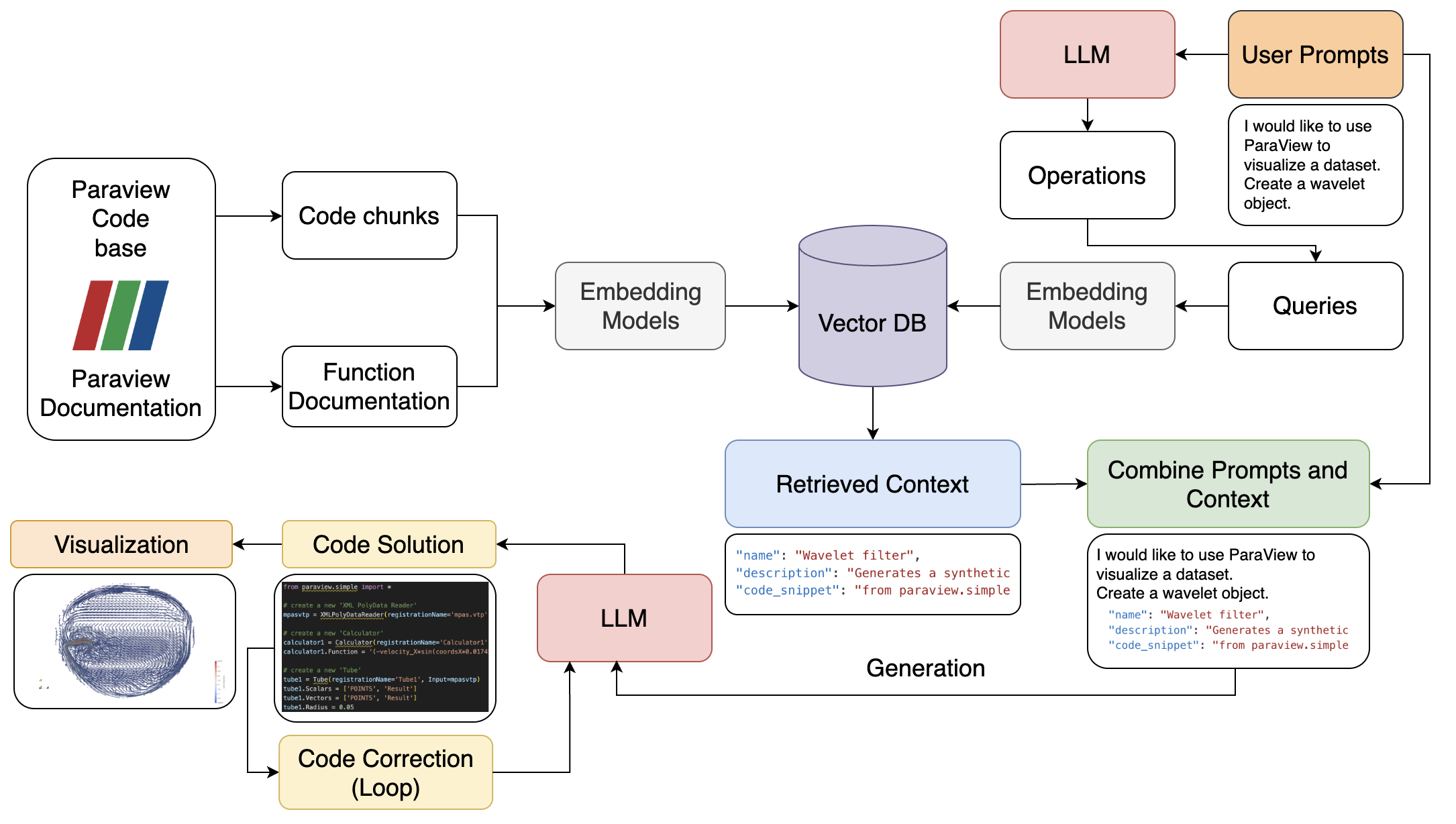}
    \caption{System architecture for LLM-assisted ParaView scripting. The process involves embedding models, a vector database, and iterative code correction.}
    \label{fig:methodology}
\end{figure*}

\subsubsection{Image Comparison Metrics}

The easiest and perhaps most common image metrics are pixel-wise comparisons. These include local differences at each pixel and global statistics such as peak signal-to-noise ratio (PSNR), root-mean-squared error (RMSE), and L-norms such as L1, L2, and L-infinity.
While easy to compute and understand, pixel-wise metrics may not match human intuition of similarity. Two images that are slightly misaligned or with slightly different color mapping would have low pixel-wise metrics but appear highly similar to a human.
For this reason, structural comparisons of image features can be more informative.

The multi-scale structural similarity metric (MS-SSIM)~\cite{wang2003multiscale} is among the most successful techniques for assessing perceptual similarity.
MS-SSIM measures differences in luminance and contrast, rather than simple pixel intensity.
However, MS-SSIM does not account for spatial distortions between images (for example, two images shifted relative to each other)~\cite{sampat2009complex}.
As a result, MS-SSIM may not accurately reflect human preferences in tests of perceptual similarity~\cite{zhang2018unreasonable}.

Learned perceptual image patch similarity (LPIPS)~\cite{zhang2018unreasonable} uses pre-trained deep networks to compute the perceptual similarity between two images by comparing their major features.
 The Berkeley Adobe Perceptual Patch Similarity (BAPPS) dataset is used to assess the quality of LPIPS as a similarity metric, and in the last 7 years has become a widespread reference dataset for perceptual image comparison~\cite{severo2023unreasonable}.
Inspired by \cite{zhang2018unreasonable}, several other perceptual metrics based on deep networks have been developed.
Bhardwaj et al. created the Perceptual Information Metric (PIM)~\cite{bhardwaj2020unsupervised}, which uses deep networks to improve a metric based on mutual information.
The Linear Autoregressive Similarity Index (LASI)~\cite{severo2023unreasonable} was developed to investigate perceptual similarity without resorting to deep features.
To compute perceptual similarity, LASI computes the pixelwise difference of the images in the embedding space and then gives the L2 norm of this difference.

\subsubsection{Video Comparison Metrics}


Cheung et al.~\cite{cheung2003efficient} propose measuring ideal video similarity as the percentage of clusters of similar frames, approximated by the volume of intersecting Voronoi cells of similar clusters. A randomized algorithm selects a small set of sampled frames for the clustering.
Seshadrinathan and Bovik~\cite{seshadrinathan2007structural} extend the SSIM structural similarity metric to video for the purpose of video quality assessment.
Similar to SSIM for static images, structural similarity often agrees with human intuition better than pixel-wise metrics such as PSNR or RMSE.
Beyond merely computing SSIM frame-by-frame, this work utilizes motion information to capture temporal artifacts.
Fuentes al.~\cite{fuentes2012similarity} compute similarity between videos using various characteristics, mainly used for human content, including speech and visual features to compute correspondences between segments and an overall similarity score.

\subsubsection{Code Comparison Metrics}

Because our task is software generation, it is natural to consider code comparison metrics as an evaluation tool. However, our initial investigation of several of the tools listed below showed that the code comparisons did not agree with image metrics, while the image metrics agreed more closely with our human intuition.
Since the purpose of the code is to produce images or animations, we decided to use those metrics over the code metrics described below.

Predating LLMs for software generation, code comparison tools have been and continue to be used in academia to detect plagiarism. For example, the Lichen plagiarism detection tool~\cite{peveler2019lichen} is used in the Submitty homework submission system in several universities and the Google Summer of Code.
The BLEU metric was originally devised to match n-grams in human language. Ren et al.~\cite{ren2020codebleu} adapted the method for code synthesis by augmenting the n-gram matches with abstract syntax trees and code semantics, resulting in CodeBLEU.
Chen et al.~\cite{chen2021evaluating} note that CodeBLEU is not always a good metric of functional correctness, and so they devised a numerically stable definition of the pass@k metric, originally defined by Kulal et at.~\cite{kulal2019spoc}. In our paper, in addition to image metrics, we use pass@k with k = 1 to record generation of an image or animation, free of syntax errors in the script, without regard to the content of the image or animation.


\section{Methodology}
\label{sec:method}

The methodology presented in Figure \ref{fig:methodology} is a structured approach to leveraging LLMs and vector databases (vector DBs) to assist users in generating and refining ParaView Python code from natural language prompts.
The approach integrates RAG with automated code correction loops to ensure accurate and efficient visualization scripting.

\begin{table*}[!t]
\caption{Benchmark examples}
\begin{tabular}{ | >{\centering\arraybackslash}p{1.25in} | >{\centering\arraybackslash}p{1.25in} | >{\centering\arraybackslash}p{1.25in} | >{\centering\arraybackslash}p{1.25in} | >{\centering\arraybackslash}p{1.25in} | }
\hline
\multicolumn{5}{| c |}{\textbf{Canonical Visualizations}} \\
\hline
\includegraphics[height=0.75in]{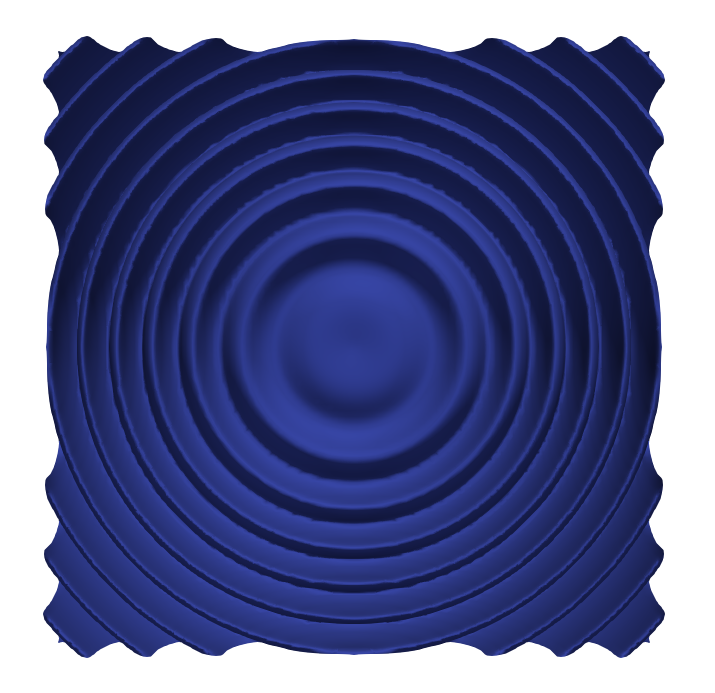}  &
\includegraphics[width=1.25in]{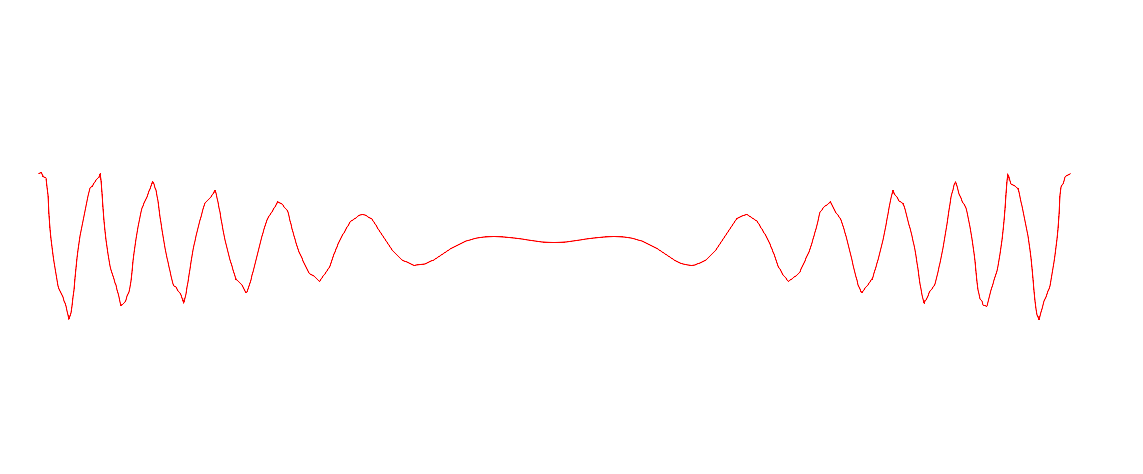} &
\includegraphics[height=0.75in]{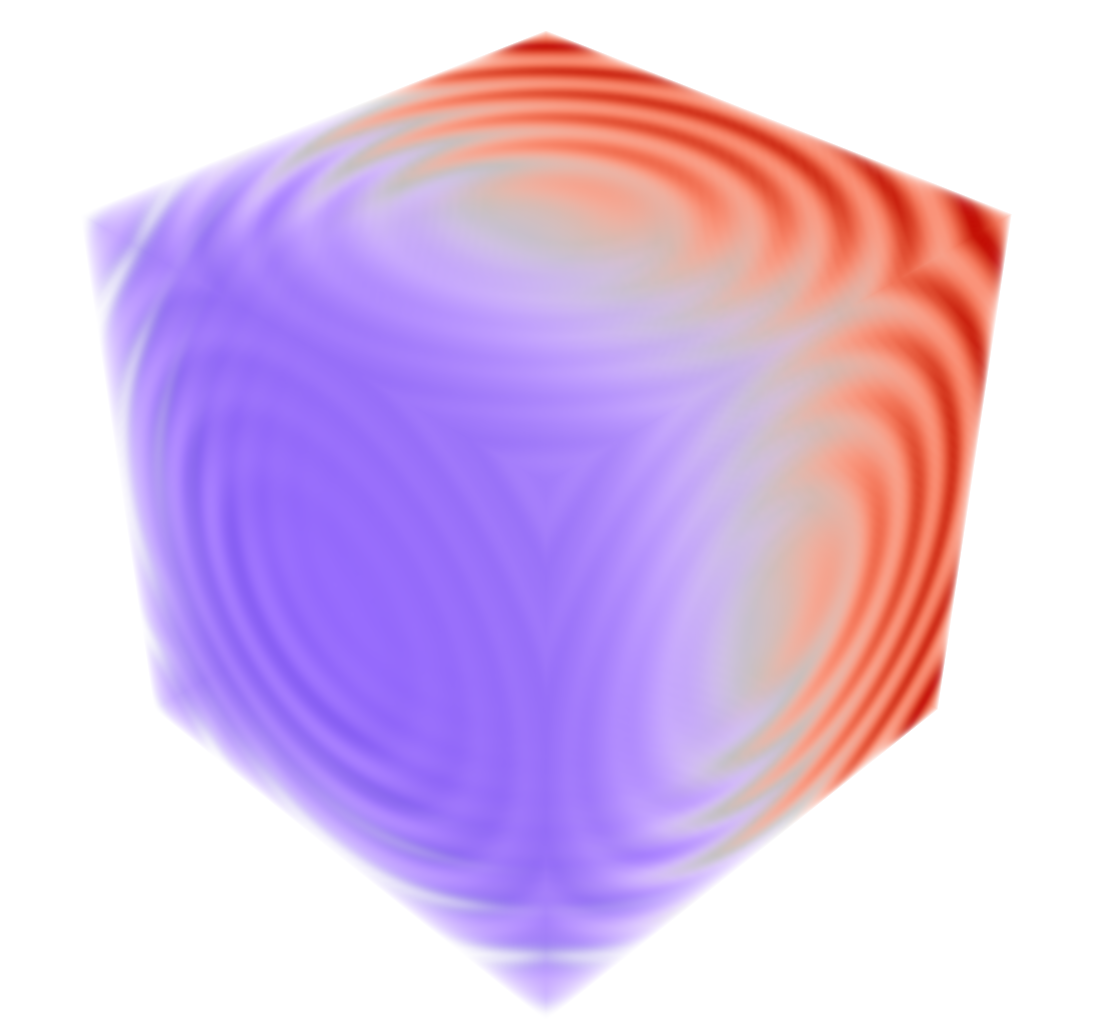}  &
{\includegraphics[height=0.75in]{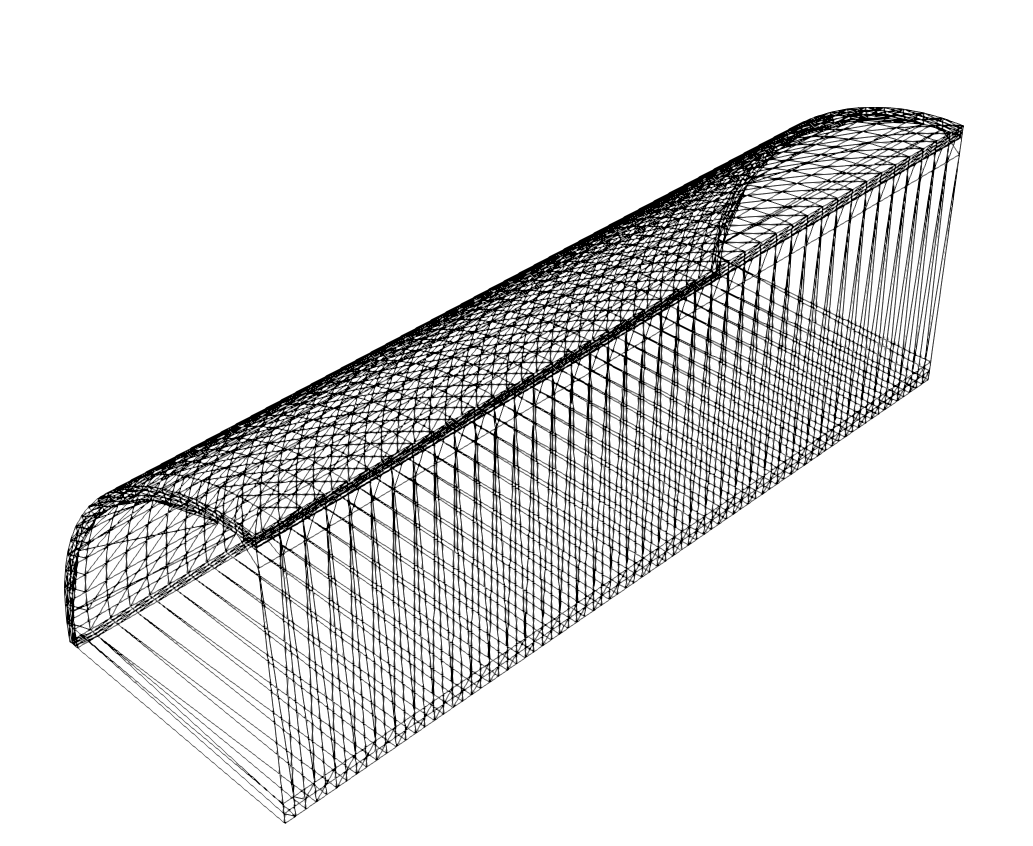}} &
\includegraphics[height=0.75in]{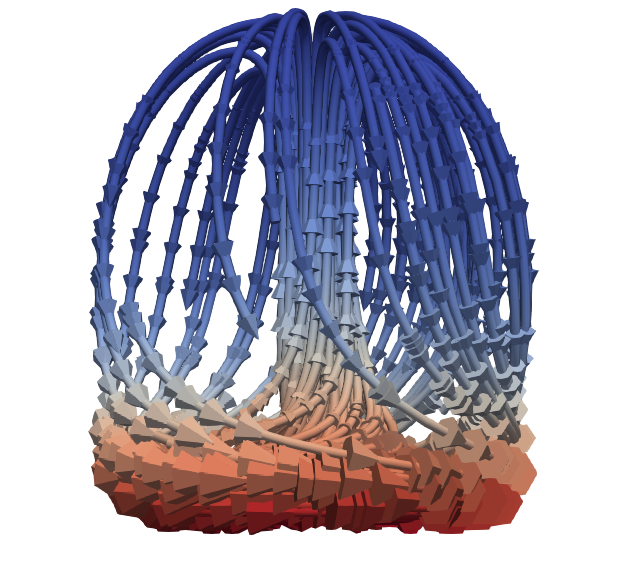} \\
ml-iso: isosurfacing & ml-slice-iso: slicing followed by contouring & ml-dvr: direct volume rendering & points-surf-clip: Delaunay triangulation followed by clipping & stream-glyph: streamline tracing \\
\hline
\includegraphics[height=0.75in]{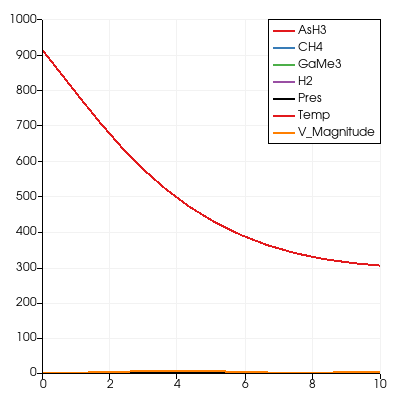} &
\includegraphics[height=0.75in]{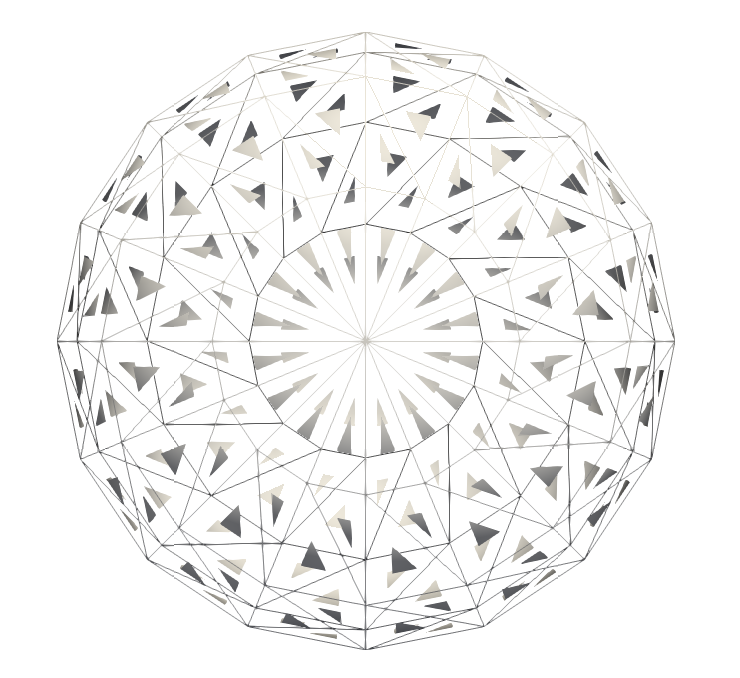} &
{\includegraphics[width=1.25in]{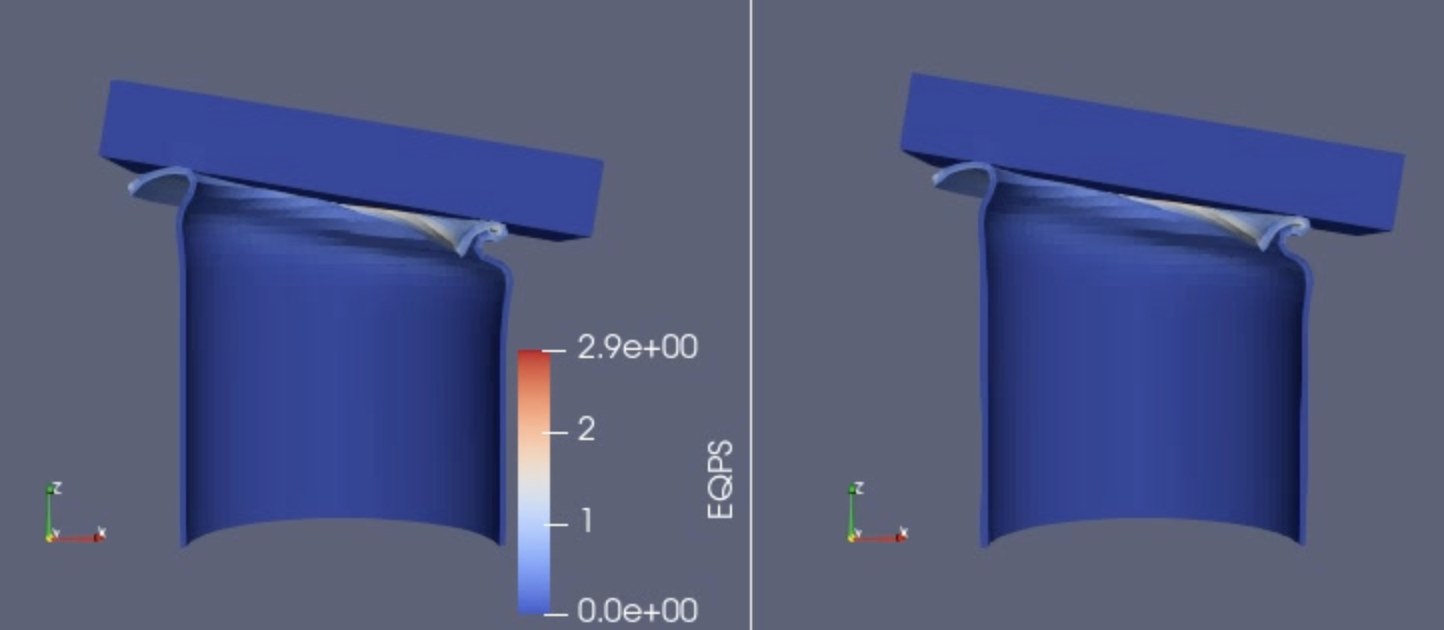}} &
\multicolumn{2}{| c |}{} \\
line-plot: chart plotting & shrink-sphere: shrink filter and wireframe rendering & time-varying: temporal interpolation and animation rendering &
\multicolumn{2}{| c |}{} \\
\hline
\multicolumn{5}{| c |}{\textbf{ParaView Regression Tests}} \\
\hline
\includegraphics[height=0.75in]{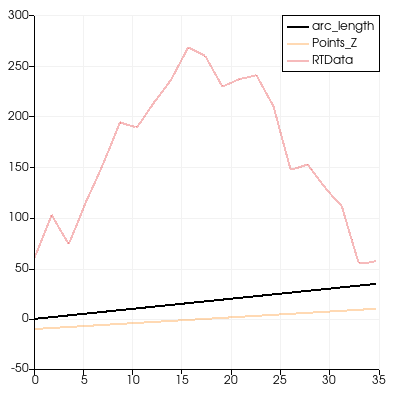}  &
\includegraphics[height=0.75in]{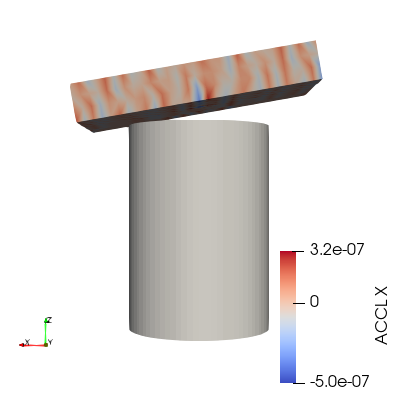} &
\includegraphics[height=0.75in]{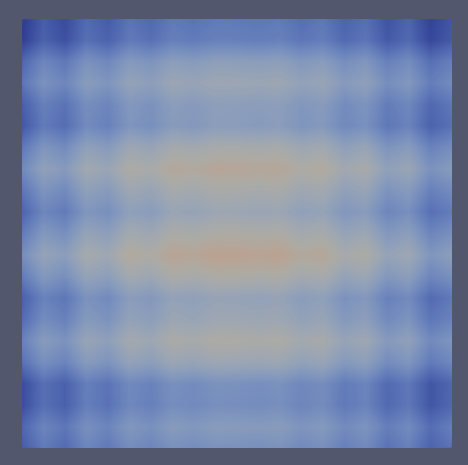}  &
\includegraphics[height=0.75in]{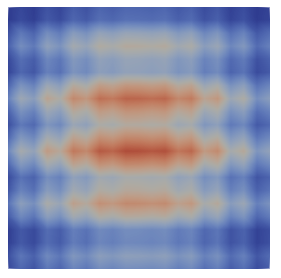}  &
\includegraphics[height=0.75in]{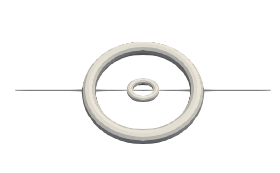} \\
chart-opacity: chart plotting with background modification & color-blocks: color mapping, VTK blocks & color-data: color mapping, calculator & export-gltf: glTF file writing & import-gltf: glTF file reading, window resizing \\
\hline
\includegraphics[height=0.75in]{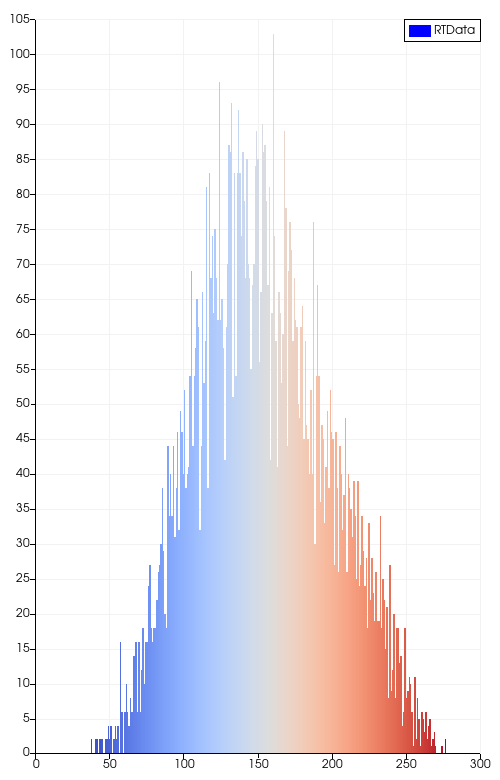} &
\includegraphics[height=0.75in]{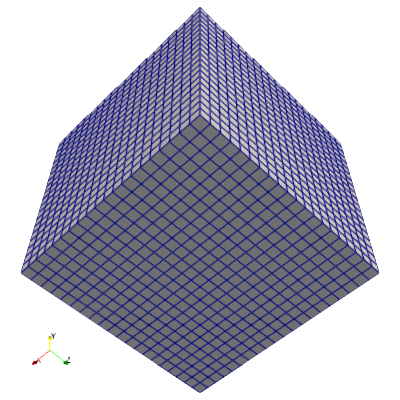} &
\includegraphics[height=0.75in]{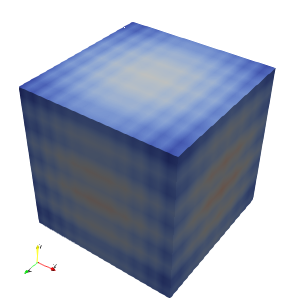} &
\includegraphics[height=0.75in]{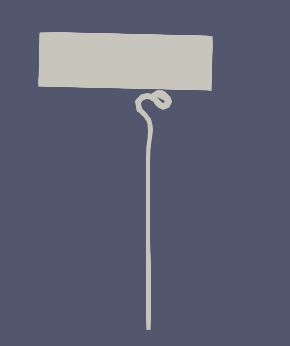} &
\includegraphics[height=0.75in]{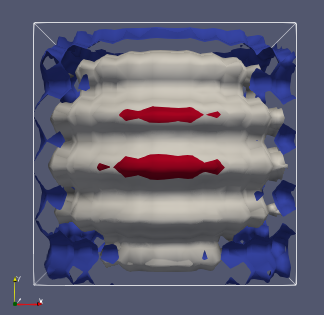} \\
render-histogram: histogram plotting  & reset-camera-direction: camera manipulation  & save-transparent: background transparency  & subseries-of-time-series: time series IO, slicing, VTK blocks  & write-ply: multiple isosurfacing, PLY file writing  \\
\hline
\multicolumn{5}{| c |}{\textbf{Scientific Use Cases}} \\
\hline
{\includegraphics[height=0.75in]{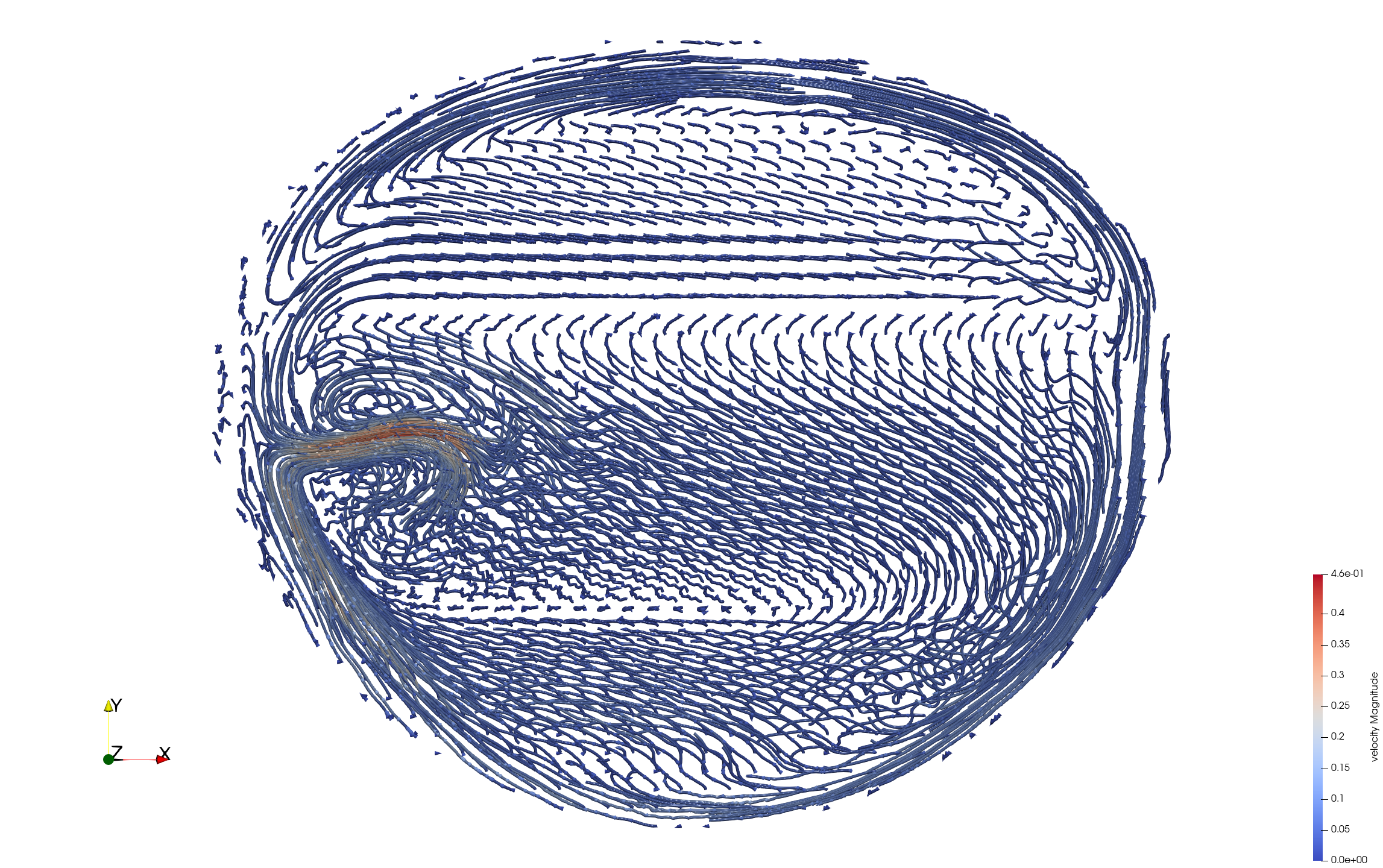}}  &
{\includegraphics[height=0.75in]{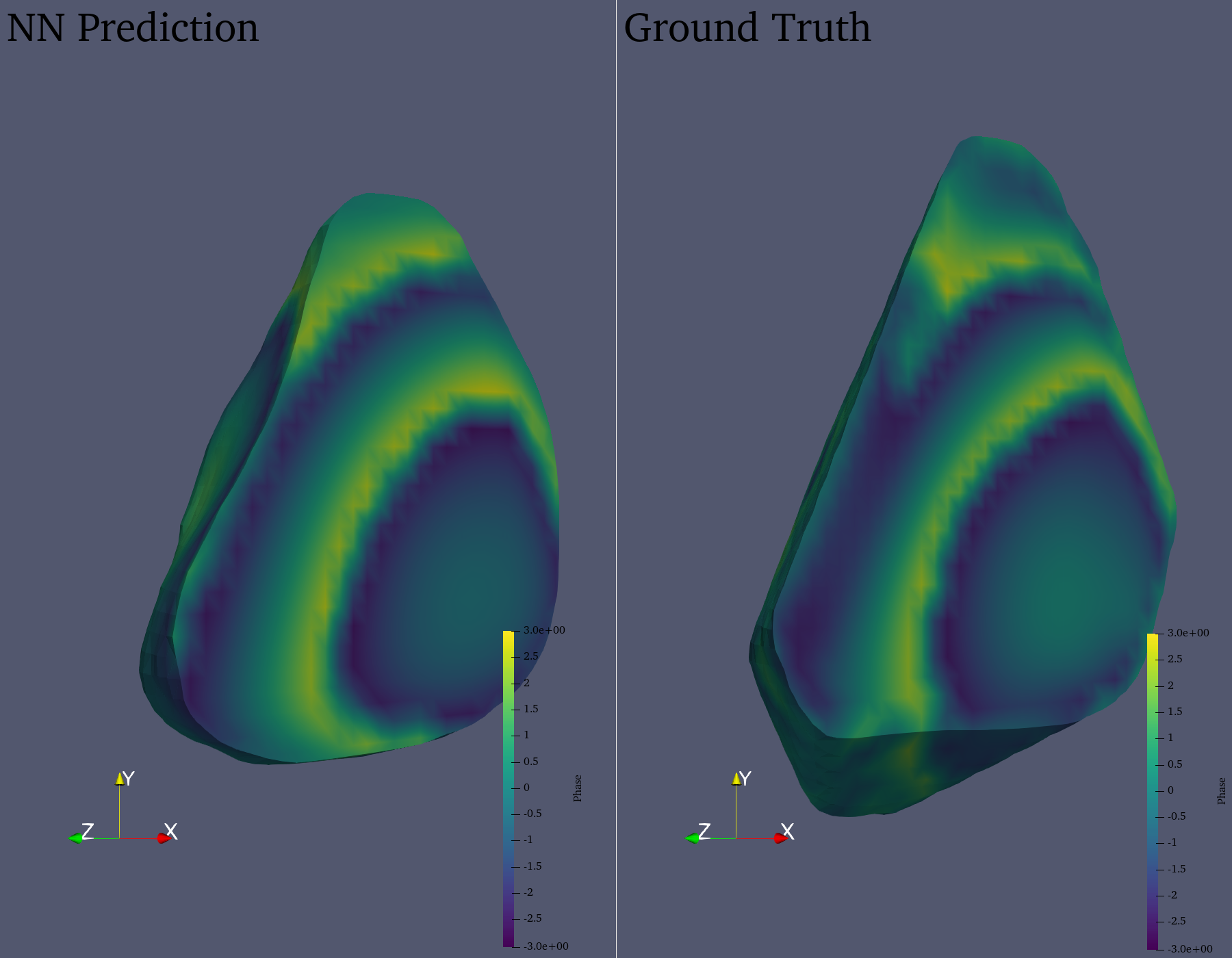}} &
\multicolumn{3}{| c |}{} \\
climate science: lat-long to geospatial, tube and glyph rendering with lighting and specularity & materials science: side-by-side comparison, isovolume, clipping, labeling, color mapping &
\multicolumn{3}{| c |}{} \\
\hline
\end{tabular}
\label{tab:examples}
\end{table*}

\subsection{Data Sources and Preprocessing}
The foundation of this methodology relies on two primary data sources: the ParaView code base and its corresponding documentation.
The code base contains various implementations and function definitions essential for performing visualization tasks, while the documentation provides detailed explanations of function usage and parameters.
To make this information accessible for automated code generation, embedding models are employed.
We use the all-MiniLM-L6-v2 embedding model from Hugging Face~\cite{huggingface24}.
These models process both code chunks and function documentation, transforming them into numerical vector representations.
These vectors are then stored in a vector DB, enabling fast and efficient retrieval of relevant information.
We use the Faiss library to implement the vector DB~\cite{douze2024faiss}.
This preprocessing step ensures that later stages can retrieve contextualized information for user queries without requiring exhaustive manual searches.

\subsection{User Interaction and Query Processing}
We initiate our methodology with  users submitting natural language prompts that outline their intended ParaView operations.
These prompts often describe complex multistep tasks, requiring multiple stages of data transformation, filtering, rendering, and visualization.
To process such complex multistep requests, we decompose the natural language prompt into structured operations that correspond to specific ParaView API functions. 

Our approach begins by parsing the user’s prompt to extract key actions such as data loading, applying filters, rendering, and output generation.
We do so by passing the user's prompt through the LLM with targeted instructions designed to explicitly identify the required operations without the need for further interpretation or description.
In so doing, we convert the user's prompt into a few short prompts that systematically break down complex natural language instructions into distinct ParaView operations, similar to chain-of-thought reasoning.
For some steps, we explicitly separate different applications of the same operation.
For example, we differentiate between the \kw{Show} command for a \kw{Tube} filter and \kw{Show} for a \kw{Glyph} filter, and we differentiate \kw{SaveScreenshot} from \kw{SaveScreenshot} with a desired background color to differentiate between prompts that specify background color and those that do not.
These expanded prompts ensure that the LLM focuses solely on extracting actionable ParaView operations, mapping each step in the user’s prompt to the corresponding ParaView API function.

The LLM generates the expanded prompts line by line, identifying the essential components necessary for execution.
By explicitly instructing the model to list each operation separately, we ensure that key steps such as loading data, performing mathematical transformations, and rendering geometric representations are distinctly listed. 
By enforcing this structured operation extraction, we create a clean, consistent, and executable sequence of steps that maps directly to ParaView’s API.
We then use the extracted list of operations to retrieve the corresponding code descriptions and code snippets from the vector DB.

\subsection{Context Retrieval and Augmentation}
To enhance accuracy and ensure precise function mapping, we convert these queries into vector embeddings using embedding models, enabling similarity-based retrieval from a vector DB containing pre-indexed ParaView documentation and example code snippets.
Once relevant documentation and code snippets are retrieved, we combine them with the user’s prompt to create a contextually enriched input for the LLM, ensuring that the generated script follows correct syntax, best practices, and logical sequencing of operations.
We ensure that essential visualization steps---such as applying filters, modifying rendering attributes, adjusting views, and saving outputs--—are correctly mapped to corresponding ParaView API calls, resulting in an executable and optimized Python script.
By systematically integrating RAG with structured prompt interpretation, we enable the accurate and reproducible generation of ParaView scripts, even for complex multistep visualization tasks.

The vector DB retrieves the relevant information, which is combined with the original user prompt to enhance context awareness.
This step is critical because user queries alone may not provide enough detail for accurate code generation.
By integrating retrieved documentation, function descriptions, and example code snippets, the LLM receives a more informed input for code generation.
This RAG approach ensures that the generated code aligns with ParaView’s API specifications and best practices.
The combination of user input and retrieved context is then formatted into a structured representation, making it suitable for subsequent processing by the LLM.

\subsection{Code Generation, Execution, and Iterative Correction}
With the enhanced prompt and relevant retrieved context, the LLM generates a structured Python script utilizing ParaView’s API.
ChatVis currently uses GPT4o as its LLM.
After generating the ParaView Python script, we execute it using ParaView’s \kw{pvpython} command to produce the intended visualization.

If the generated Python code fails to execute, further refinement is required.
To handle this, we developed a tool that extracts error messages from the \kw{pvpython} output.
This tool systematically processes output logs, identifying potential execution issues and enabling an iterative refinement cycle.

The tool begins by breaking the output into individual lines and creating a list to store detected messages.
It scans for tracebacks, recognized by the presence of the keyword \kw{File}, and continues collecting relevant lines until it detects specific error types like \kw{AttributeError}.
After gathering all necessary information, the function compiles the extracted lines into a structured list and returns the error messages for further processing.

Once errors are identified, the extracted messages are sent to the LLM, along with a contextual prompt instructing it to correct the script.
The LLM analyzes the error messages and modifies the script accordingly to resolve the issue.
The revised script is then executed again, repeating the process of error detection, extraction, and correction until an error-free version is achieved.
This approach significantly enhances the usability and reliability of the system, allowing seamless script generation and execution with minimal human intervention.

Once the final version of the script is generated and refined through the error detection and correction loop, we execute it once more using \kw{pvpython} to produce the visualization.
To validate the accuracy of the generated script, we capture a screenshot of the visualization output.
This visual output serves as a reference for assessing whether the script meets the intended requirements and produces the expected results.
The next section describes a set of experiments using a suite of test cases to evaluate the resulting accuracy of ChatVis and several unassisted LLMs.


\section{Experiments and Results}
\label{sec:results}

Experiments are conducted using a suite of 20 benchmark example tasks described below. Metrics of performance include percentage of tasks that executed without syntax errors (pass@1), and average scores for SSIM, PSNR, and LPIPS, each averaged over the tasks that were error-free.
Performance for several currently available top-performing  LLMs, unassisted by ChatVis, is collected.
To further evaluate the utility of our vector DB and RAG methodology, we compare performance for ChatVis with and without RAG, where the latter method only includes few-shot prompting and iterative error correction, but no access to ParaView documentation or code snippets.
To evaluate the sensitivity to prompt wording, we also compare ChatVis performance for two versions of user prompts.

\begin{table}[b!]
    \caption{Effectiveness of ChatVis and various state-of-the-art LLMs in performing visualization tasks.}
    \label{tab:results-overall}
          \begin{tabular}{| p{0.15\textwidth} | p{0.05\textwidth}  | p{0.05\textwidth} | p{0.05\textwidth}  | p{0.05\textwidth} |}
	   \hline
       \textbf{LLM} & \textbf{pass@1} $\uparrow$  & \textbf{SSIM} $\uparrow$ & \textbf{PSNR} $\uparrow$ & \textbf{LPIPS} $\downarrow$ \\
	   \hline
	   ChatVis  & \textbf{95} & \textbf{0.80} & \textbf{40.1} & \textbf{0.26}  \\ 
	   o1pro  & 50 & 0.70 & 35.7 & 0.45 \\
	   o3-mini-high  & 35 & 0.58 & 30.7 & 0.55 \\
	   GPT-4.5-Preview & 40 & 0.66 & 32.8 & 0.43 \\
	   Claude-3.7-Sonnet  & 40 & 0.68 & 34.4 & 0.44 \\
	   LLaMa-3.1-405B  & 15 & 0.57 & 24.7 & 0.59 \\
	   \hline
   \end{tabular}
\end{table}

\subsection{Benchmark Example Tasks}
We developed an evaluation benchmark suite of 20 visualization tasks of varying complexity.
Each task consists of a reference image, a manually-written Python script that produces that image, and two natural language prompts of differing verbosity that describe the desired visualization.
One prompt is written with detailed instructions and contains more visualization jargon that might appear in the ParaView documentation; we refer to this as the ``full prompt.''
The second prompt is written in a slightly relaxed conversational style and contains less (but still some) visualization jargon; we refer to this as the ``quick prompt.''

The 20 tasks in our benchmark, depicted in Table~\ref{tab:examples}, are grouped into three categories of roughly increasing complexity. These are:

\paragraph{Canonical Visualizations.} Eight tasks consist of relatively well-known visualization operations like volume rendering, isocontouring, clipping, and line plots.
Each of these tasks require only a small number of operations.
In addition, they do not include more technical requests like reading and writing in different file formats, or changing image resolution.

\paragraph{ParaView Regression Tests.} Ten tasks in the benchmark are adapted from ParaView's own regression testing framework.
These tasks exercise several (but by no means all) of ParaView's visualization routines.
They may also combine several different operations in a row.
They include more instructions that relate to a dataset's representation, like reading/writing data in different file formats, or interfacing with the concept of VTK ``blocks.''

\paragraph{Scientific Use Cases.} Two tasks in the benchmark are drawn from real visualization workflows used by scientists.
One use case from the field of climate science visualizes a dataset describing ocean currents.
This task involves transforming data between coordinate systems, visualizing vectors with stream tubes and glyphs, and customizing lighting effects.
The second use case, from materials science, visualizes a nanocrystal reconstructed from a diffraction pattern captured by a high-energy x-ray synchrotron light source.
In this task, a side-by-side view is created to compare two reconstructions.
The data are mapped from cells to vertices; an isovolume is computed, and then the isovolume is clipped, colored, and labeled.

When creating the natural language prompts for our benchmark, we considered the tradeoff between specificity and user-realism.
A chat assistant for any task should be expected to behave reasonably even when instructions from the user are incomplete or vague.
Since users typically interact with chat models in a conversational way, it would be unrealistic to create prompts that are excessively detailed and exhaustive in their description of the desired visualization.
For example, our prompts rarely specify characteristics like lighting or render window size; we expect ChatVis to simply use Paraview's default settings or to make reasonable default assumptions when user instructions are incomplete.
The prompts also leave certain settings unspecified that could have multiple default values that a user would find satisfactory. Background color,~\footnote{%
  Some prompts do include explicit instructions regarding background color and opacity to test ChatVis's ability to control these settings when asked.} for instance, is rarely specified in the prompts, but could be set to any number of colors that a user might find acceptable.
Similarly, positioning the camera can typically be done in multiple ways to produce a reasonable visualization.
In order to create a benchmark that accurately represents expected user behavior, our prompts allow for variability in the generated visualization.

\begin{figure}[t!]
\begin{subfigure}{0.16\textwidth}
\includegraphics[width=\linewidth]{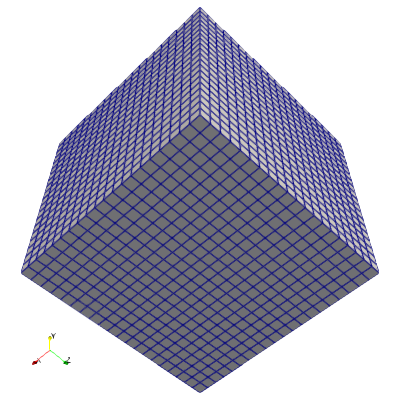} 
\caption{Ground truth}
   \label{fig:rcd-original}
   \end{subfigure}
\begin{subfigure}{0.16\textwidth}
\includegraphics[width=\linewidth]{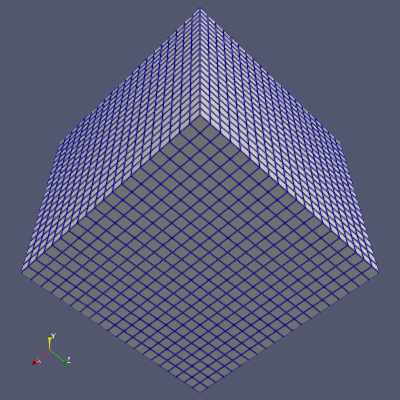}
\caption{ChatVis}
 \label{fig:rcd-chatvis}
 \end{subfigure}
\begin{subfigure}{0.16\textwidth}
\includegraphics[width=\linewidth]{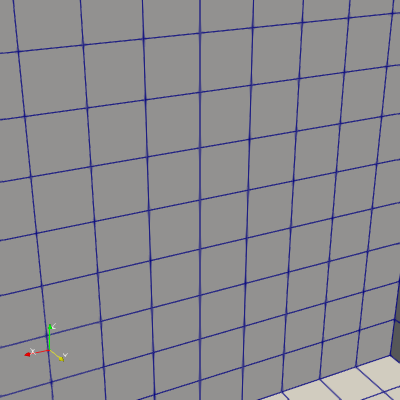}
\caption{o1pro}
 \label{fig:rcd-o1pro}
 \end{subfigure}
 \caption{Generated images for reset-camera-direction task.} 
 \label{fig:rcd}
\end{figure}

\subsection{Comparisons}

For visualization tasks, we compare ChatVis against the state-of-the-art large language models, both open- and closed-source. We selected the following models because they are among the top-performing LLMs available at the time of writing.

\begin{itemize}

\item \textbf{o1pro and o3-mini-high} are closed-source language models from OpenAI with strong reasoning capabilities that provide step-by-step reasoning.

\item \textbf{GPT-4.5-Preview} is the latest GPT model from OpenAI with its improved capability to follow user instructions in various tasks such as coding and writing. 

\item \textbf{Claude-3.7-Sonnet} is a closed-source language model developed by Anthropic designed as a hybrid reasoning model for high-level problem-solving and coding tasks. 

\item \textbf{LLaMA-3.1-405B}~\cite{touvron2023llama} is an open-source language model from Meta with 405 billion parameters. LLaMA-3.1-405B enhances code generation, text summarization, and natural language understanding compared with previous models. 

\end{itemize}

\paragraph{Metrics.}

The script generated by ChatVis and the other LLM models is executed using the \kw{pvpython} command of ParaView version 5.13.1 to produce the visualization.
 To validate the accuracy of the generated script, we capture a screenshot of the visualization output. This visual output serves as a reference for assessing whether the script meets the intended requirements and produces the expected results.

We evaluate the effectiveness of LLMs in performing visualization tasks in terms of both syntax correctness and image quality metrics.
For syntax correctness, we use pass@1, which indicates whether the generated Python script can run without any syntax errors, regardless of its output image quality.
For image quality metrics, we use SSIM, PSNR, and LPIPS when comparing generated images by LLMs against the ground truth.
SSIM ranges from 0.0 to 1.0, with higher score indicating better quality.
PSNR increases from 0.0 with improving image quality, ie. higher is better.
LPIPS decreases from 1.0 to 0.0 with improving quality, ie., lower is better.

The ground truth image is generated by the reference Python script that we include with each visualization task in the benchmark.
We report the average scores over the  visualization tasks that executed free of syntax errors.
For example, if an LLM successfully completed only 15\% of the tasks, then PSNR, SSIM, and LPIPS are averaged over only 3 out of 20 tasks.
One could alternatively scale the image metrics over all tasks by multiplying by the fraction of tasks completed; see Equation~\ref{eq:scaling}.
~
\begin{equation}
\label{eq:scaling}
\begin{split}
	PSNR_{scaled} = pass@1/100 * PSNR \\
	SSIM_{scaled} = pass@1/100 * SSIM \\
	LPIPS_{scaled} = 1.0 - (1.0 - LPIPS) * pass@1 / 100
\end{split}
\end{equation}

\subsection{Performance of ChatVis}

We first evaluate ChatVis in performing scientific visualization by comparing it against the state-of-the-art LLMs.
Table~\ref{tab:results-overall} summarizes the obtained correctness and image quality metrics for the benchmark suite of visualization tasks.
We observe that ChatVis outperforms the state-of-the-art LLMs in both syntax correctness and image quality metrics.
In particular, we see a large difference in pass@1 scores for syntax correctness. For example, ChatVis successfully completes 95\% of the visualization tasks, while o1pro---the best performing model among the unassisted LLMs---is only able to complete half of the tasks.
Moreover, we see that quality of the generated images is much higher with ChatVis.

To further highlight the differences in performing scientific visualization tasks, we show the generated images for the reset-camera-direction task in Figure~\ref{fig:rcd}.
Figure~\ref{fig:rcd-original} is generated from the hand-written script in our benchmark suite and represents the ground truth, while Figure~\ref{fig:rcd-chatvis} is generated by ChatVis, and Figure~\ref{fig:rcd-o1pro} is generated by o1pro.
We observe that ChatVis can correctly perform the requested operations.
The background color is not specified in the prompt, leading to different background colors.
On the other hand, o1pro is not able to reset the camera direction correctly, capturing only a portion of the dataset instead of the intended view.

To illustrate the differences in the generated Python scripts, we report the generated scripts created by ChatVis and o1pro for another example, the color-blocks task.
The prompt provided by the user is shown below for this task.

\begin{myenv}{User prompt}
I would like to use ParaView to visualize a dataset.
Please generate a ParaView Python script for the following operations.
Set the background to a blue-gray palette.
Read the file 'can.ex2'.
This is a multiblock dataset.
Color the dataset by the vtkBlockColors field.
Retrieve the color map for vtkBlockColors.
Retrieve the opacity transfer function for vtkBlockColors.
Retrieve the 2D transfer function for vtkBlockColors.
Set block coloring for the block at $/IOSS/element_blocks/block_2$ using the variable ACCL on the x component of the points.
Rescale the block's color and opacity maps to match the current data range of block\_2.
Retrieve the color transfer function for the ACCL variable of block\_2.
Enable the color bar for block\_2.
Apply a cool to warm color preset to the color map for block\_2.
Set the camera to look down the -y direction and to see the entire dataset.
Save a screenshot of the visualization in the file 'color-blocks-screenshot.png'.
\end{myenv}

\setbox0=\hbox{%
\begin{minipage}{.48\textwidth}
\begin{lstlisting}[
basicstyle={\tiny\ttfamily},
identifierstyle={\color{black}},
tabsize=2,
language={python},
numbersep=8pt,
numbers=left,
xleftmargin=0.5cm,frame=tlbr,framesep=2pt,framerule=0pt
]               
from paraview.simple import *

# Load the BlueGrayBackground palette for rendering
LoadPalette('BlueGrayBackground')

# Read the multiblock dataset file 'can.ex2'
canex2 = OpenDataFile('can.ex2')

# Get or create the render view and reset the camera
renderView1 = GetActiveViewOrCreate('RenderView')
renderView1.ResetCamera(False, 0.9)  # Ensuring the data fits well

# Display the dataset with surface representation
canex2Display = Show(canex2, renderView1, 'UnstructuredGridRepresentation')
canex2Display.Representation = 'Surface'


# Apply scalar coloring using the 'vtkBlockColors' field
ColorBy(canex2Display, ('FIELD', 'vtkBlockColors'))

# Retrieve color, opacity, and 2D transfer functions for vtkBlockColors
vtkBlockColorsLUT = GetColorTransferFunction('vtkBlockColors')
vtkBlockColorsPWF = GetOpacityTransferFunction('vtkBlockColors')
vtkBlockColorsTF2D = GetTransferFunction2D('vtkBlockColors')

# Set block scalar coloring 
ColorBlocksBy
(canex2Display, ['/IOSS/element_blocks/block_2'], ('POINTS', 'ACCL', 'X'))

# Rescale block color and opacity transfer functions
canex2Display.RescaleBlocksTransferFunctionToDataRange
(['/IOSS/element_blocks/block_2'], False, True)

# Retrieve the color transfer function for 'ACCL' in 'block_2'
blockACCLLUT = GetBlockColorTransferFunction
('/IOSS/element_blocks/block_2', 'ACCL')

# Enable the scalar bar (color legend) for 'block_2'
canex2Display.SetBlocksScalarBarVisibility
(renderView1, ['/IOSS/element_blocks/block_2'], True)

# Apply the 'Cool to Warm' color preset to the color map for 'block_2'
blockACCLLUT.ApplyPreset('Cool to Warm', True)

# Reset the camera to look in the -y direction and ensure the dataset is in view
renderView1.ResetActiveCameraToNegativeY()

# Render the scene (necessary after camera or color changes)
Render()

# Save a screenshot of the visualization
SaveScreenshot('color-blocks-screenshot.png', renderView1)

\end{lstlisting}
\end{minipage}
}
\savestack{\listingA}{\box0}
\setbox0=\hbox{%
\begin{minipage}{.48\textwidth}
\begin{lstlisting}[
basicstyle={\tiny\ttfamily},
identifierstyle={\color{black}},
tabsize=2,
language={python},
numbersep=8pt,
numbers=left,
moredelim={[is][\textcolor{red}]{\^}{\^}}, 
xleftmargin=0.5cm,frame=tlbr,framesep=2pt,framerule=0pt
] 
from paraview.simple import *

# 1) Use a blue-gray palette for the background and interface
^LoadPalette(paletteName="Blue Gray Background")  # Wrong palette name^

# 2) Read the file 'can.ex2'
canReader = ExodusIIReader(FileName=["can.ex2"])
canReader.UpdatePipeline()

# Create a render view and show the data
renderView = CreateView('RenderView')
renderView.ViewSize = [1200, 800]
# Show the dataset
canDisplay = Show(canReader, renderView)

# 3) Color the dataset by the vtkBlockColors field 
ColorBy(canDisplay, ('FIELD', 'vtkBlockColors'))

# 4) Retrieve the color map, opacity transfer function, and 2D transfer function
#    for "vtkBlockColors"
vtkBlockColorsLUT = GetColorTransferFunction('vtkBlockColors')
vtkBlockColorsPWF = GetOpacityTransferFunction('vtkBlockColors')
vtkBlockColorsTF2D = GetTransferFunction2D('vtkBlockColors')

# 5) Set block coloring for the block 
^canDisplay.SetBlockColorArrayName 
("/IOSS/element_blocks/block_2", ('POINTS', 'ACCL')) # Not a valid function call^
^canDisplay.SetBlockColorArrayComponent 
("/IOSS/element_blocks/block_2", 0)  # Not a valid function call^

# 6) Rescale the block's color and opacity maps 
canDisplay.RescaleTransferFunctionToDataRange
(False, True, "/IOSS/element_blocks/block_2") ^ # Wrong order of arguments^

# 7) Retrieve the color transfer function for the ACCL variable of block_2
acclLUT = GetColorTransferFunction('ACCL')

# 8) Enable the color bar for block_2
^canDisplay.SetBlockColorLegendVisibility
("/IOSS/element_blocks/block_2", True) # Not a valid function call^

# 9) Apply a "Cool to Warm" color preset to the color map for block_2
acclLUT.ApplyPreset("Cool to Warm", True)

# 10) Set the camera to look down the -y direction and see the entire dataset
# First let ParaView reset the camera to fit all data
renderView.ResetCamera()

# Manually set camera direction to look along -y: 
camera = GetActiveCamera()

# We'll move the camera to negative y, pointing toward the center:
focalPoint = [0, 0, 0]
camera.SetFocalPoint(focalPoint)
camera.SetPosition(0, -1, 0)   # far in -y
camera.SetViewUp(0, 0, 1)      # Z up
# Expand the camera clipping range so everything is visible
^ResetCameraClippingRange() # Not a valid function call^

# 11) Save a screenshot of the visualization
SaveScreenshot("color-blocks-screenshot.png", 
               renderView, ImageResolution=[1200, 800])

\end{lstlisting}    
\end{minipage}
}
\savestack{\listingB}{\box0}

\begin{table*}[t]
\centering
\begin{tabular}{|c|c|}
\hline
{\listingA} &
{\listingB} \\
\hline
\end{tabular}
\vspace{-0.1in}
\caption{Generated Python scripts with ChatVis (left), and o1pro (right) for the color-blocks task.}
\vspace{-0.1in}
\label{tab:script-cb}
\end{table*}

We show the generated scripts in Table~\ref{tab:script-cb}.
We can see that o1pro generates hallucinations in several places, due to the lack of knowledge about this specific visualization task.
For example, the script encounters an error when attempting to set block coloring, calling nonexistent function calls (\kw{SetBlockColorArrayName} and \kw{SetBlockColorArrayComponent} at lines 26-29).
Moreover, when rescaling the block colors, o1pro uses a correct function call, but this time it incorrectly orders the arguments (line 33).
Further hallucinations occur when o1pro uses nonexistent function calls to enable the color bar for the block (line 39) and expand the camera clipping range (line 58).
In contrast, ChatVis successfully executes all these operations by utilizing the correct function calls.

\subsection{Role of RAG}

To highlight the effectiveness of retrieval augmented generation (RAG), we compare it against the few-shot prompting approach without RAG, which only corrects syntax errors.
Few-shot prompting is what a user would do manually when interacting with an LLM, continually prompting the LLM to correct errors but otherwise providing no other external data from a vector DB.
Table~\ref{tab:chatvis-approach} shows the obtained correctness and image quality metrics corresponding to these different approaches for the benchmark suite of visualization tasks.
With RAG, ChatVis completes 95\% of the tasks, while with few-shot prompting we complete 70\%.
Recall from Table~\ref{tab:results-overall} that the best unassisted LLM completed 50\% of the tasks in the benchmark.

We limit the few-shot prompting approach to a maximum of 5 iterations of error correction.
We found that without RAG, typically 2-5 iterations of error correction were executed, while with RAG, generally only one iteration was needed because the vector DB supplied ChatVis with the correct function calls.
An additional advantage of RAG over few-shot prompting is scalability.
Each iteration to correct syntax errors grows the inference context by the length of the generated code and the list of errors encountered, with the potential to exceed the maximum context length of the LLM eventually.
RAG does not have this limitation.

\begin{table}[h!]
\vspace{-0.1in}
    \caption{Comparison of few-shot prompting vs. retrieval augmented generation for visualization tasks.}
    \label{tab:chatvis-approach}
       \begin{tabular}{| p{0.15\textwidth} | p{0.05\textwidth}  | p{0.05\textwidth} | p{0.05\textwidth}  | p{0.05\textwidth} |}
	   \hline
       \textbf{Approach} & \textbf{pass@1} $\uparrow$  & \textbf{SSIM} $\uparrow$ & \textbf{PSNR} $\uparrow$ & \textbf{LPIPS} $\downarrow$ \\
	   \hline
	   RAG  & \textbf{95} & \textbf{0.80} & \textbf{40.1}& \textbf{0.26}  \\ 
	   Few-shot prompting  & 70 & 0.69 & 34 & 0.46  \\ 
	   \hline
   \end{tabular}
\end{table}
\vspace{-0.15in}
\begin{table}[h!]
\vspace{-0.15in}
    \caption{Completeness and image quality metrics for the ChatVis under different  prompting strategies.}
    \label{tab:prompt-sensitivity}
          \begin{tabular}{| p{0.15\textwidth} | p{0.05\textwidth}  | p{0.05\textwidth} | p{0.05\textwidth}  | p{0.05\textwidth} |}
	   \hline
       \textbf{Strategy} & \textbf{pass@1} $\uparrow$  & \textbf{SSIM} $\uparrow$ & \textbf{PSNR} $\uparrow$ & \textbf{LPIPS} $\downarrow$ \\
	   \hline
	   Full prompt  & 95 & 0.80 & 40.1 & 0.26  \\ 
	   Quick prompt  & 75 & 0.78 & 39.6 & 0.33 \\
	   \hline
   \end{tabular}
\end{table}
\vspace{-0.25in}
\subsection{Prompt Sensitivity}
\label{sec:prompt-sensitivity}
\vspace{-0.07in}
We evaluate the sensitivity of our approach with two different prompting strategies---full prompt and quick prompt---that vary in length.
Below are the two prompts for the climate science use case, for example.
The performance of ChatVis under these different prompting strategies appears in Table~\ref{tab:prompt-sensitivity}.
The results show that ChatVis is affected by prompt length, with its syntax completion rate decreasing from 95\% to 75\% while maintaining similar image quality when using shorter prompts.
The observed drop in completion rate is expected, as shorter prompts may omit important details, as previously illustrated.
However, breaking down the scientific visualization problem step-by-step---a la chain-of-thought reasoning---along with iterative error correction enables ChatVis to still produce better results than unassisted LLMs, even with less input from the user.

\vspace{-0.07in}

\begin{myenv}{Full prompt}
I would like to use ParaView to visualize a dataset of ocean currents.
Please generate a ParaView Python script for the following operations.
Read in the file named 'mpas.vtp'.
Apply a calculator filter to compute the following function:

$(-velocity_X*sin(coordsX*0.0174533) + velocity_Y*cos(coordsX*0.0174533)) * iHat + (-velocity_X * sin(coordsY*0.0174533) * cos(coordsX*0.0174533) - velocity_Y * sin(coordsY*0.0174533) * sin(coordsX*0.0174533) + velocity_Z * cos(coordsY*0.0174533)) * jHat + 0*kHat$

Render the computed values using a tube filter with 0.05 as the tube radius.
Color the tubes by the magnitude of the velocity.
Light the tubes with the maximum shininess and include normals in the lighting.
Add cone glyphs to show the direction of the velocity.
The glyphs are composed of 10 polygons, having a radius 0 0.15, a height of 0.5, and a scaling factor of 0.5.
View the result in the -z direction.
Adjust the view so that the tubes occupy 90\% of the image.
Save a screenshot of the result in the filename 'soma-screenshot.png'.
The rendered view and saved screenshot should be 2294 x 1440 pixels with a white background.
\end{myenv}

\begin{myenv}{Quick prompt}
Please generate a ParaView Python script for the following operations.
Read in the file named 'mpas.vtp'.
Apply a calculator filter to convert velocity from geospatial to lat-long coordinates.
Render the computed values using a tube filter with 0.05 radius, colored by velocity magnitude, and lit with maximum shininess, and include normals for lighting.
Add cone glyphs to show the direction of the velocity, using 10 polygons, radius 0.15, height 0.5, and scaling factor 0.5.
View the result in the -z direction scaled so that the tubes occupy most of the image.
Save a screenshot of the result, 2294 x 1440 pixels, white background, in the filename 'soma-screenshot.png'.
\end{myenv}


\section{Science Use Cases}
\label{sec:science}

We highlight two science use cases because they are the most complex examples in our benchmark.
In both cases, Chatvis produced a high-quality result, whereas none of the unassisted models could produce an executable script.
Table~\ref{tab:science-results} shows the image metrics for the full prompt version of both science cases.

\begin{table}[b!]
    \caption{Image quality metrics for ChatVis science use cases with full prompt.}
    \label{tab:science-results}
          \begin{tabular}{| p{0.15\textwidth} | p{0.05\textwidth}  | p{0.05\textwidth} | p{0.05\textwidth}  |}
	   \hline
       \textbf{Science Case} & \textbf{SSIM} $\uparrow$ & \textbf{PSNR} $\uparrow$ & \textbf{LPIPS} $\downarrow$ \\
	   \hline
	   Climate science  & 0.99 & 36.2 & 0.01  \\ 
	   Materials science & 0.97 & 46.8 & 0.05 \\
	   \hline
   \end{tabular}
\end{table}

\subsection{Climate Science}

MPAS-Ocean is the ocean simulation component of the U.S. Department of Energy's (DOE) collection of climate science codes collectively known as E3SM~\cite{e3sm-model}.
The Simulating Ocean Mesoscale Activity (SOMA)~\cite{wolfram2015diagnosing} test case for MPAS-Ocean consists of a double-gyre large eddy system in a hypothetical region in the ocean.
The SOMA system is often used to test simulation parameters and to execute Lagrangian particle tracing to generate pathlines of ocean currents over long periods of time.
In a current DOE-sponsored project, the SOMA test case is being used to develop and deploy in situ particle tracing, connecting external particle tracing codes with MPAS-Ocean through an online workflow.
The pathline trajectories are stored in VTK format and visualized offline using ParaView.

\begin{figure}[t!]
    \centering
	\includegraphics[width=2.5in]{figs/soma-screenshot.png}
	\includegraphics[width=2.5in]{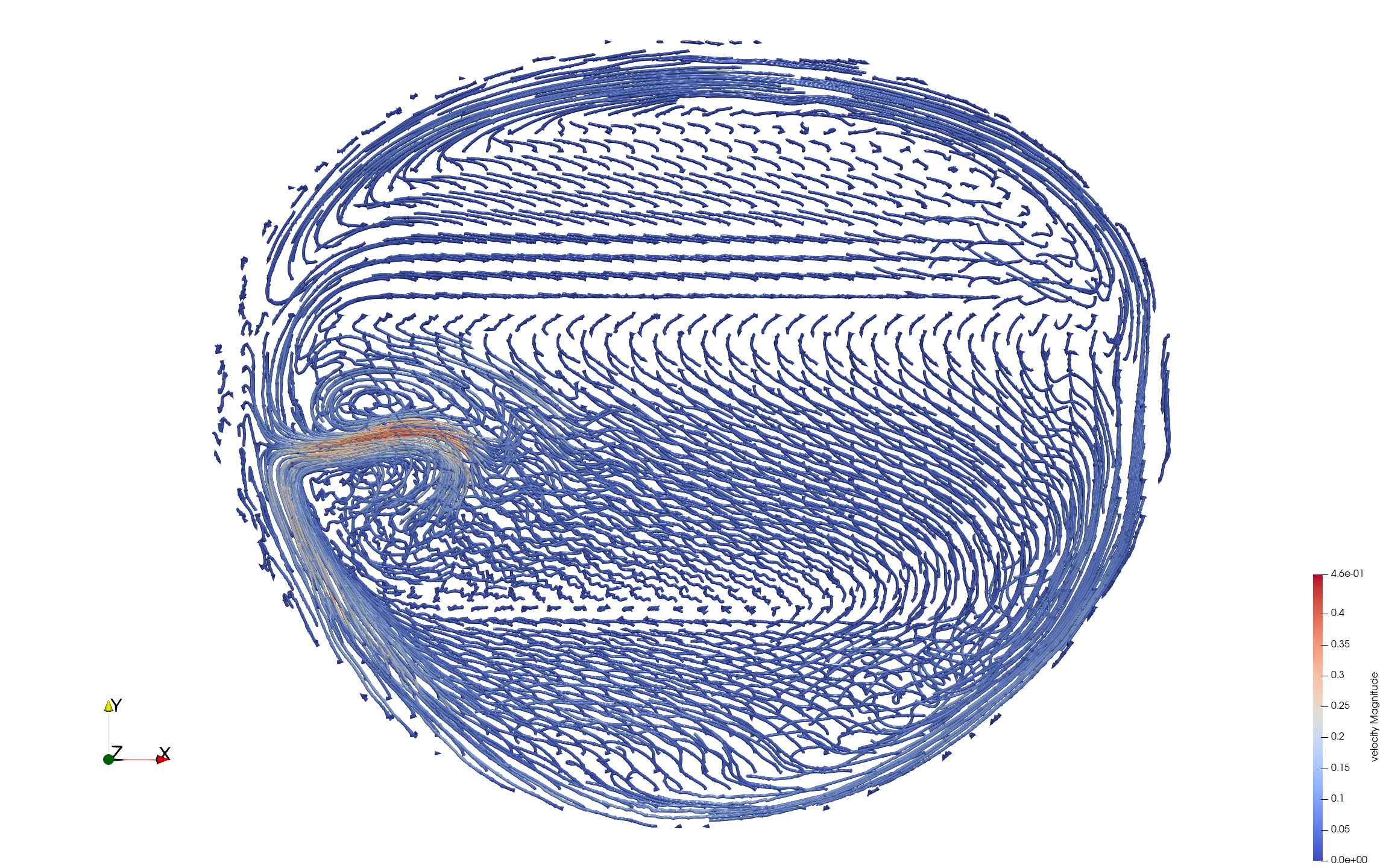}
    \caption{Climate science: pathlines traced from a vector field of the SOMA test case of the MPAS-Ocean simulation. Top: screenshot generated by ChatVis. Bottom: ground-truth reference image.}
    \label{fig:climate}
\end{figure}

The visualization is the result of executing the 32 km resolution SOMA test case running MPAS-Ocean for two years of simulation time, outputting data every 2 days, coupled in situ with a particle tracing code from the FTK~\cite{guo_tvcg21} package, tracing 2500 pathlines initiated on a grid of 50 x 50 seeds.
Executing ChatVis on the full prompt featured in Section~\ref{sec:prompt-sensitivity}, compared with the ground-truth reference image resulting from the hand-written script, is shown in Figure~\ref{fig:climate}.
Although the images are visually very similar, we note a slight discrepancy in the data range of the color bar between the two images: the ground truth color bar range is [0.0, 0.46] while the ChatVis-generated color bar range is [0.0, 0.34].
Not exactly an error, the reason for this is ambiguity in the prompt and resulting scripts.
The order of the glyph and tube representations is swapped between the generated script and ground truth scripts.
Because there is one color legend, the scale of the last operation takes precedence, hence the difference in color bar ranges. 
Since the prompt does not specify the order of operations, both results are ``correct'' with respect to the prompt.
A more explicit prompt would resolve the ambiguity.

\subsection{Materials Science}
Researchers at Argonne National Laboratory's Advanced Photon Source are working to automatically invert 3D coherent diffraction images (CDI) rapidly and accurately using AI models.
This approach is in contrast to the traditional method of reconstructing CDI images, which involves a computationally expensive, iterative process to retrieve the phase component of the complex-valued diffraction pattern, given that detectors at the instrument only capture the magnitude of the light waves, and not their phase.
The Autophase-NN project~\cite{yao2021autophasenn} couples coherent diffraction data generation with training and inference of unsupervised physics-aware deep learning models.
The performance of deep learning models can be continuously improved with data from new coherent imaging experiments, but our understanding  and trust of the AI workflow relies on careful visual comparison with the iterative solution.

\begin{figure}[h!]
    \centering
	\includegraphics[width=2.0in]{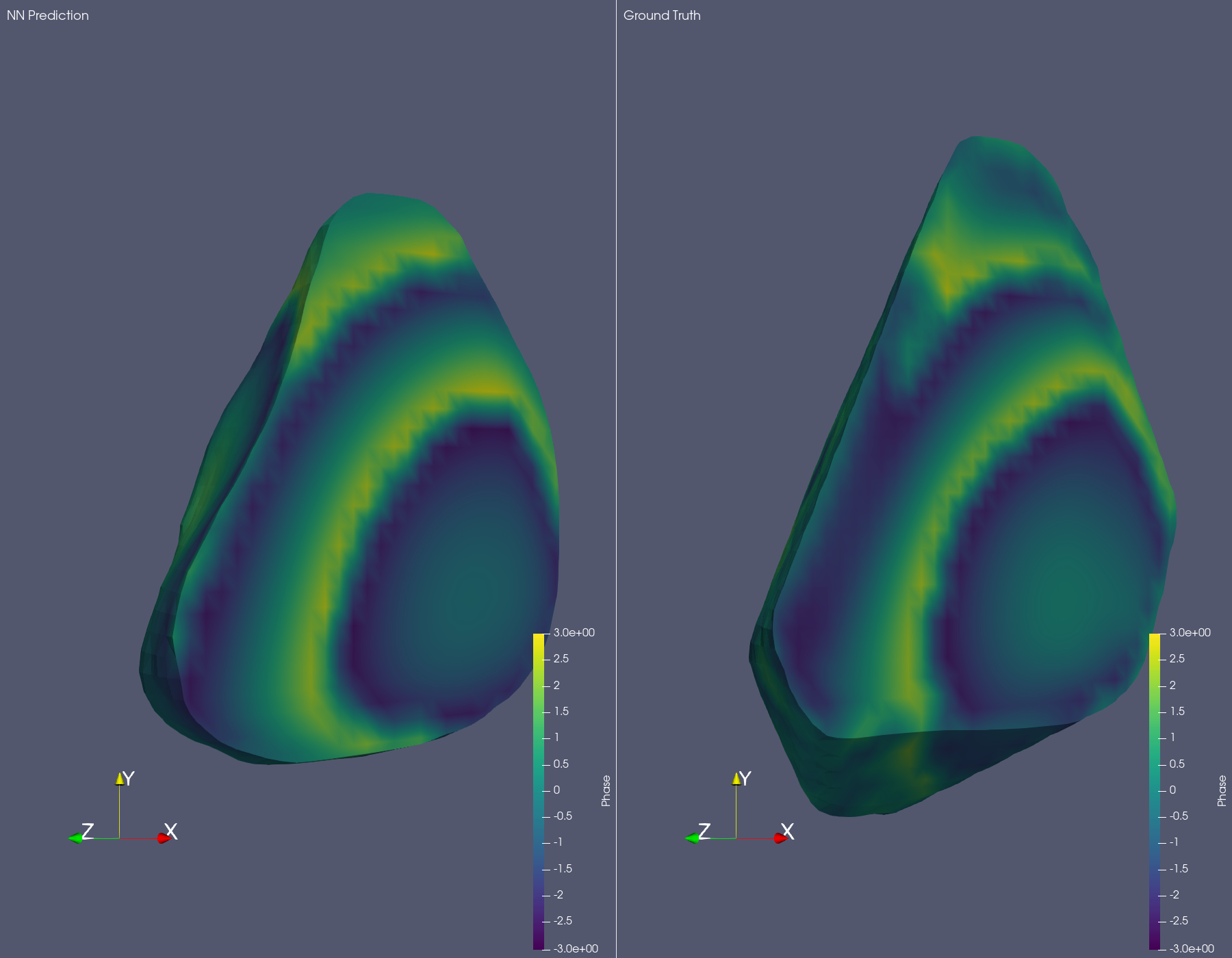}
	\includegraphics[width=2.0in]{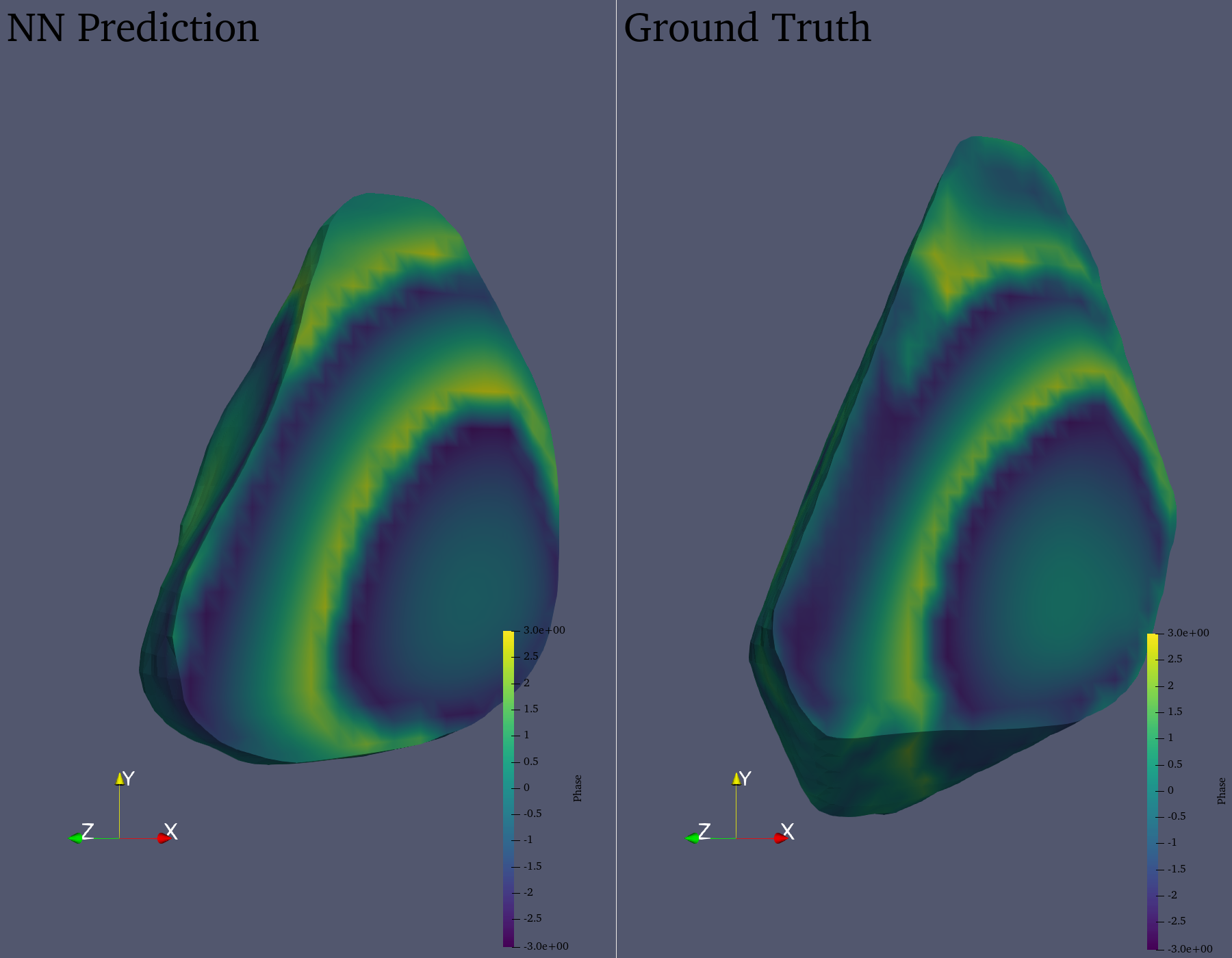}
    \caption{Materials science: comparison of coherent diffraction imaging of a nanocrystal in a high-energy synchrotron light source reconstructed using a traditional iterative method compared with using a neural network Top: screenshot generated by ChatVis. Bottom: ground truth reference image.}
    \label{fig:materials}
\end{figure}

Figure~\ref{fig:materials} shows one such comparison for the reconstruction of a nanocrystal, visualized by slicing an isovolume.
Comparing the result generated by ChatVis with the reference image from a hand-written script, we see good correspondence.
Slight differences exist in the color and font of the text in the title and color bar, but similar to the climate science case, these differences are due to ambiguity in the prompt, which appears below.

\begin{myenv}{Full prompt}
Please generate a ParaView Python script for the following operations.
I want to compare two datasets side by side.
Create two views each of size 900 pixels wide x 1400 pixels high.
In the left view, read the dataset in the file "neg10\_prediction.vtr".
In the right view, read the dataset in the file "neg10\_ground\_truth.vtr".
In both views, convert the "Intensity" and "Phase" variables from cell to point data.
In both views, take an isovolume of the "Intensity" variable in the range of [0.2, 1.0].
In both views, clip the data with a clipping plane located at (32.0, 32.0, 32.0) and +x normal direction.
In both views, use the Viridis (matplotlib) color map for the "Phase" variable, scaled to the data range.
Add a colormap legend in both views.
Label the left view "NN Prediction".
Label the right view "Ground Truth".
Orient the camera to look in the (-1, 0, -1) direction.
Position the camera so that the dataset fits in the view.
Save a screenshot of both views in the file "comparison-screenshot.png".
\end{myenv}

\section{Conclusion}
\label{sec:conclusion}


\subsection{Summary}

We presented an assistant, ChatVis, that aids an LLM to generate Python code for ParaView scientific visualization tasks, without the need for retraining or fine-tuning the LLM.
ChatVis employs chain-of-thought prompt simplification, retrieval-augmented prompt generation using a vector database of documentation and code examples, and error checking with iterative feedback to correct errors.

In order to evaluate ChatVis, we developed a benchmark suite of canonical visualization tasks, ParaView regression tests, and scientific use cases.
We used pass@1, PSNR, SSIM, and LPIPS to evaluate and score the resulting images.
We found that ChatVis outperforms all the other unassisted models that we tested by significant margins. Compared with the best-performing standalone LLM (o1pro) in our experiments, ChatVis generated nearly twice as many scripts without syntax errors, improved PSNR by nearly 5 dB, improved SSIM by approximately 10\%, and improved LIPS by nearly a factor of two. Those image metrics are averaged over only the completed examples; scaling the metrics by the number of error-free scripts generated would show even greater improvement for ChatVis.

We also measured the effect of using a vector DB and RAG compared with only few-shot prompting.
We found the performance of ChatVis to be enhanced by RAG through the vector DB.
Compared with few-shot prompting alone, RAG with the vector DB improved the pass@1 rate by 25\%, and image metrics similarly improved.
We also measured the effect of prompt verbosity by using two versions---full prompt and quick prompt---of user prompts.
We found as expected that condensing the user prompt reduced the number of completed examples by 20\%, and the image metrics averaged over the completed examples worsened by small amounts. Even with the quick prompt versions, however, ChatVis easily outperformed the best unassisted models that were given full prompts.

Although the unassisted models generated Python scripts that often were close to being correct, they usually had a few syntax errors, calling hallucinated or outdated API commands or using incorrect arguments.
In other cases, even when the script did not have syntax errors, an incorrect image was generated.
For 3d and 4d scientific visualization, code that is nearly correct is not sufficient, because finding and fixing the remaining bugs requires the same expert knowledge as to generate a new script from scratch.

\subsection{Ongoing and Future Work}

We envision several directions for continuing this research.
Expanding the benchmark to include tasks employing more ParaView filters and additional science use cases would make the benchmark more comprehensive.
For the time-varying examples, we evaluated results using image comparison metrics on the final frame; another approach would be to employ video comparison metrics on the entire animation.
Exploration of the wording of the user prompt could be broadened to include sensitivity to variation in tone, quality of information content, clarity, and style.
In this paper we compared our implementation using RAG and a vector DB with few-shot prompting. Another option would be to fine-tune one of the open-source LLMs and compare with RAG and few-shot prompting.

We believe a similar approach as ChatVis could be applied to other scientific visualization packages, namely VisIt.
We plan to release ChatVis as open-source software in the coming months so that others can apply ChatVis to their own visualizations. We welcome contributions from the visualization community to the project. We also plan to release our benchmark suite of test cases as a standalone package separately from ChatVis so that it can be used to evaluate other approaches.


\acknowledgments{
This material is based upon work supported by the U.S. Department of Energy, Office of Science, Office of Advanced Scientific Computing Research, under contract numbers DE-AC02-06CH11357, program manager Margaret Lentz.
The authors gratefully acknowledge the assistance of Luke Van Roekel of Los Alamos National Laboratory, Hanqi Guo of the Ohio State University, and of Mathew Cherukara of the Argonne Advanced Photon Source in compiling the science use cases in this paper.
}

\bibliographystyle{abbrv-doi-hyperref}

\bibliography{paper}

\end{document}